\documentstyle[aps,epsf,prl]{revtex}
\def\simle{\mathrel{{}^<_\sim}}

\begin{document}
\twocolumn[\hsize\textwidth\columnwidth\hsize\csname
@twocolumnfalse\endcsname
\draft
\title{A Limit on Primordial Small-Scale Magnetic
Fields from CMB Distortions}

\author{Karsten Jedamzik}
\address{Max-Planck-Institut f\"ur Astrophysik}
\address{85748 Garching bei M\"unchen, Germany}
\author{Vi\v{s}nja Katalini\'c and Angela V. Olinto}
\address{Department of Astronomy and Astrophysics and Enrico 
Fermi Institute}
\address{University of Chicago, 5640 S. Ellis Ave, Chicago, IL 60637}

\maketitle

\begin{abstract}
Spatially varying primordial magnetic
fields may be efficiently dissipated prior to the epoch of recombination due to
the large viscosity of the baryon-photon fluid. We show that this
dissipation may result in observable chemical potential $\mu$ and
Compton $y$
distortions in the cosmic microwave background (CMB)
spectrum. Current upper limits on $\mu$ and $y$ from FIRAS constrain magnetic
fields to have strength $B_0 < 3\times 10^{-8}\,$Gauss (scaled to
the present) between comoving coherence length $\approx 400$ pc and 
$\approx 0.6$ Mpc. These represent the strongest upper
limits on small-scale primordial magnetic fields to date.
\end{abstract}
\pacs{PACS numbers: 98.80.Cq, 95.85.Sz, 98.58.Ay, 98.70.Vc}
\vskip2.2pc]


At present not much is known about the existence or absence of magnetic
fields in the early universe. There is a plethora of proposed 
scenarios for the generation of primordial magnetic
fields~\cite{H83,fieldcreation,V91,EO94}.   Most of these scenarios
operate by \lq\lq causal\rq\rq\ mechanisms,  such that
the coherence length
of the resulting magnetic field is usually limited by the horizon
scale at the epoch of magnetogenesis (e.g., $\sim 1\,$pc
at the QCD transition), unless magnetic field energy density is
subsequently transferred
from smaller to larger scales by inverse cascades~\cite{cascade}. 

 Limits on the total magnetic field energy density
extant at the epoch of Big Bang nucleosynthesis (BBN) may be derived 
(i.e. $B_0 \,{}^<_{\sim}\, 10^{-6}\,$Gauss) from the possible increase of
 $^4$He production beyond observational bounds~\cite{BBN}. 
(Here, and henceforth, we quote magnetic field strengths at the
values they would have at the present epoch, i.e. after
cosmological expansion.) Nevertheless, such strong magnetic fields
usually do not survive neutrino induced dissipation 
prior to BBN~\cite{JKO98}, rendering this limit useless in most
cases. Fields on large scales may be limited by the cosmic microwave
background (CMB) and by Faraday rotation measures of light from high
redshift quasars. A homogeneous cosmic magnetic field on the scale of the
present horizon is limited by its effect on the CMB due to an induced
anisotropic cosmic expansion to fall below 
$B_0  \simle 3\times 10^{-9}\,$Gauss~\cite{BFS97} while the Faraday
rotation limits are slightly weaker at $B_0  \simle,  
10^{-9}\,$Gauss~\cite{Fara}. For smaller coherence lengths, Faraday
rotation measures constrain primordial fields with a 50 Mpc coherence
length at present to have $B_{50 {\rm Mpc}}
\simle 6 \times 10^{-9}\,$Gauss  and 
$B_{\rm Mpc} \simle  10^{-8}\,$Gauss for Mpc coherence lengths~\cite{Fara}.
In addition, a spectrum dependent limit on stochastic magnetic fields
of $B_0\simle  10^{-7}\,$Gauss on $\sim 0.1$ Mpc scales has been quoted
due to the possible CMB distortions from magnetic field generated 
gravitational waves~\cite{Pen95DFK99}.

Future high-precision CMB
satellite observations should be able to detect (or limit)  large-scale
magnetic fields.  Observational proposals include the detection of
rotational (vector) perturbations below~\cite{SB98}  and
above~\cite{DKY98} the Silk damping scale, a depolarization of the cosmic
microwave background by Faraday rotation~\cite{HHZ97},  as well as 
distortions of the acoustic peaks due to a changed (magneto-) sonic
propagation velocity at the epoch of  recombination~\cite{ADGR96}.  These
effects may be observable for magnetic field strengths above $B_0
\,{}^>_{\sim}\, 3\times 10^{-9} - 10^{-7}\,$Gauss given anticipated
sensitivities of the satellite missions, and under the assumption that
magnetic field coherence lengths are not much smaller than the comoving
Silk damping scale ($\sim 10$Mpc).

In this letter, we impose a new and fairly generic limit
on pre-recombination era magnetic
fields on small scales based on stringent limits on the deviation of
the CMB spectrum from a perfect blackbody. 
 
We have recently shown that spatially varying, stochastic, magnetic
fields undergo significant dissipation before
recombination beyond that of magnetic diffusion~\cite{JKO98}. 
To see how fields dissipate, imagine initial conditions of a
randomly magnetized baryon-photon
fluid with zero peculiar flows and at uniform fluid density. 
The non-force-free component of the field accelerates the fluid
resulting in the conversion of magnetic energy into kinetic
energy of the fluid. 
For all length scales,
$\lambda$, with magnetic relaxation time, $\tau_r \sim \lambda /v_A$
shorter than the Hubble time,
$\tau_r < H^{-1}$ (where $H$ is the Hubble parameter), the resulting
fluid motions may be described by a superposition of Alfv\'en- and slow
magnetosonic- waves with approximate energy equipartition between magnetic
and kinetic energy. Here $v_A$ is the Alfv\'en velocity, which 
well before recombination (and on 
scales larger than the photon mean free path) is
given by $v_A = B / \sqrt{ 4 \pi (\rho_{\gamma} + p_{\gamma})} = 
3.8 \times 10^{-4} c\,
(B_0 / 10^{-9} {\rm G})$, where
$\rho_{\gamma}$ and $p_{\gamma}$ are photon- energy density 
and pressure, respectively.
(In what follows we adopt natural units with the speed of light $c=1$.)
The generated kinetic energy is efficiently
dissipated into heat by the considerable shear viscosity
of the pre-recombination fluid, $\eta$, which 
arises from the long photon mean free path. The
dissipation of energy drawn from the initial magnetic field 
resumes until the magnetic field reaches a force-free state, i.e., does
not exert any magnetic stresses on the fluid. For most models of magnetic
field generation in the early universe, if an  initially random magnetic
field  relaxes to a final force-free field configuration  (for example,
straight field lines), the final state will only have a negligible
fraction of the initial magnetic field energy. 

A detailed calculation of the dissipation times for Alfv\'en, slow- and
fast- magnetosonic perturbations on a uniform magnetic field
background as a function of mode wavelength $\lambda$
was performed in~\cite{JKO98}. 
Similar to the damping of baryon-photon sound waves by photon- shear
viscosity and heat conductivity (i.e., Silk damping), effective damping of
magneto-hydrodynamic perturbations  occurs in the photon diffusion
limit, when the photon mean free path $l_{\gamma}\ll\lambda$. At a given
cosmic time, $t$, modes with $\lambda$ smaller than the photon diffusion
length, $d_{\gamma}\sim (l_{\gamma}t)^{1/2}$, are efficiently dissipated.
It should be noted that although the dissipation scales have
been computed in~\cite{JKO98} in the linear regime, it
seems difficult to avoid dissipation by non-linear effects
unless power on small scales  can be efficiently transferred  to large
scales  increasing the magnetic field coherence length  faster than
the increase in the dissipation scale. Note also that a subset of the full
non-linear MHD equations exhibits the same  damping as linear
modes~\cite{SB98_2}.  

In general, damping of Alfv\'en and slow magnetosonic waves  is
significantly different from damping of sound and fast magnetosonic waves
when $v_A\ll 1$,
corresponding to present magnetic field strengths $B_0\ll 3\times
10^{-6}\,$Gauss. When the magnetic field relaxation time, $\tau_r \sim
\lambda /v_A$, is longer than the dissipation time scale (by photon
diffusion), $\tau_d \sim \lambda^2 /l_{\gamma}$, the magnetic field
configuration becomes overdamped. In this limit,  modes take longer
than $\tau_r$ to accelerate the fluid since the fluid is strongly \lq\lq
dragged\rq\rq . For some modes, this overdamping occurs  while still in
the photon diffusion limit ($l_{\gamma}\ll\lambda$). Such modes
do not dissipate until later when the photons are
free-streaming on scales, $l_{\gamma}\gg\lambda$. In this case, the  
effective damping scale is much shorter than the photon diffusion length,
$\sim v_Ad_{\gamma}$ (cf.~\cite{JKO98}).

In this letter, we are mostly concerned with damping of
magnetic fields in the photon diffusion limit, since this leads to the
strongest constraints on small scales.
In~\cite{JKO98} the dissipation times
were computed in the WKB approximation, i.e., with mode
amplitudes evolving as $\tilde{b}_k(t) = \tilde{b}_k(t_0)\, {\rm
exp}(i\int_{t_0}^t\omega{\rm d}t)$, where $\omega$ is mode
frequency and $k$ is comoving wavevector. 
In this expression $\tilde{b}$ denotes a \lq\lq comoving\rq\rq\
magnetic field strength, scaled to the present epoch, with
$\tilde{b} = ba^2$, where $b$ is the redshift dependent magnetic field and $a$
is the scale factor ($a=1$ at the present epoch). Dissipation
of magnetic fields manifests itself through an imaginary part in the
mode frequency. The dispersion relations for Alfv\'en and slow
magnetosonic waves in the photon diffusion limit are
identical to lowest order in $b$ 
\begin{equation}
\label{dispersion}
\omega_{\rm osc}^{\rm SM,A}=v_A{\rm cos}\theta\biggl({k\over a}\biggr)
+{3\over 2}i{\eta'\over (1+R)}\biggl({k\over a}\biggr)^2\, .
\end{equation}
In this expression $3(\rho_{\gamma} + p_{\gamma})\eta' =\eta$, where
$\eta$ is shear viscosity, $\rho_{\gamma}$ and $p_{\gamma}$ as above, and  
$R=3\rho_b /(4\rho_{\gamma})\approx 0$ at redshifts $z\gg 10^4$, 
with $\rho_b$ baryon density. 
Photon shear viscosity is given by
$\eta = (4/15)\rho_{\gamma} l_{\gamma}$
with the photon mean free path
$l_{\gamma} = 1/(n_e\sigma_{Th})$ determined by Thomson scattering
on electrons.

One may Fourier transform the stochastic magnetic field by writing,
$ \tilde{b}(x) = (V^{1/2}/(2\pi k_c)^{3/2})$ $\int {\rm d}^3
k\,\tilde{b}_k\, {\rm exp}(ikx)$, where $V$ and $k_c$ are the
normalization volume and wavevector, respectively. The total
energy density of the field $\rho_{B}^0$ at the present
epoch, and in the absence of dissipation, 
is then given by spatial average, $\langle ...\rangle$,
which yields $\rho_B^0 =\langle\tilde{b}^2\rangle/ (8\pi) = 
1/(8\pi k_c^{3})\int {\rm d}^3 k\,|\tilde{b}_k|^2$. Most primordial
magnetic field models generate fields that are statistically isotropic
and can be described by a power law spectrum,
$|\tilde{b}_k|^2 = B_0^2 (k/k_c)^n (n+3)/4\pi$ for $k<k_c$ and zero
otherwise, normalized such that
$\langle\tilde{b}^2\rangle = B_0^2$ and $l_c = 2 \pi/ k_c$ corresponds to
the field coherence length.

It is well known that injection of energy into the CMB 
between redhifts $10^6 - 10^7$ and recombination,
$z\sim 10^3$, may yield
observable distortions in the CMB blackbody spectrum, generically
characterized by a chemical potential $\mu$ and a Compton $y$
parameter~\cite{distort}.
Given current upper limits on these parameters, i.e. $|\mu | < 9\times
10^{-5}$ and $y < 1.5\times 10^{-5}$,
one may infer that no more than  
$\sim 6\times 10^{-5}$ 
of the CMB energy density may have been generated 
in this redshift regime by nonthermal processes~\cite{FIRAS}.
Equilibration of photons to a blackbody distribution has to proceed
via photon-electron scattering, i.e. Compton scattering and
double-Compton scattering, slow processes
due to the smallness of the baryon-to-photon ratio. 
Here, we are mainly concerned with
blackbody distortions resulting from diffusion (or mixing) of photons
from different regions with different peculiar velocities.
This process initially yields a spectrum which may be described
by a superposition of blackbodies of different
temperature, i.e. a Compton $y$ distortions~\cite{ZIS72}.
When the mixing occurs at high redshift $z\gg 4\times 10^4$ the
optical depth of photons towards Compton scattering is still large,
redistribution of photons in frequency space is possible,
significantly diminishing the amplitude of the final $y$-distortion.
Nevertheless, for redshifts $z\, ^{<}_{\sim}\, 2\times 10^6$
the final distribution will not
contain the same photon number as a blackbody at the same temperature
$T$, due to the inefficiency of photon number creating processes
at large frequencies, of which
double-Compton scattering is the most rapid one. 
The resultant distribution will then be of the 
Bose-Einstein type at large frequencies with non-vanishing 
chemical potential $\mu$. 
In contrast, when mixing occurs at lower redshifts 
$z\simle 4\times 10^4$ spectral distortions are simply
described by a non-vanishing
Compton $y$ parameter.
For a more detailed discussion 
on the CMB equilibration
process we refer the reader to Refs.~\cite{ZIS72} and~\cite{HS93}, 
which have discussed generation of
$\mu$ and $y$ distortions resulting from primordial turbulence, very
similar to the processes under consideration here. 

It has been recently suggested~\cite{PP99} that the presence of
pre-recombination magnetic fields affects the emission of gyroradiation
by electrons into the CMB. However,
the authors erroneously argued that this emission leads to large
chemical potential distortions, even for very small magnetic fields,
due to the efficiency of creating \lq\lq extra\rq\rq\ photon number
at the fundamental, and small, frequency of gyration,
$h\nu_g /T \approx 10^{-6} (B_0/10^{-8}{\rm Gauss})(z/10^6)$. 
Unfortunately, the authors
did not consider the inverse processes, i.e., the absorption of
gyroradiation
and inverse bremstrahlung, which even for very low redshifts are 
more rapid than the Hubble expansion due to the low frequency they
occur at~\cite{distort}. 
In spite of the emission of gyroradiation, the spectrum
at low frequency remains Planckian due to the inverse reactions.  

We may now proceed to calculate the chemical potential $\mu$
distortions
resulting from magnetic field dissipation at high redshift.
The evolution of chemical potential distortions at large frequencies 
may be well approximated by~\cite{HS93}
\begin{equation}
\label{dmu}
{d\mu\over dt} = -{\mu\over t_{DC}(z)} + 1.4\, {Q_B\over
\rho_{\gamma}}\, ,
\end{equation}
where $Q_B=d\rho_B /dt$ is the instantaneous energy dissipation rate of
the magnetic field. In this expression  
$t_{DC} = 2.06\times 10^{33}{\rm
s}\, (1-Y_p/2)^{-1}(\Omega_bh^2)^{-1}z^{-9/2}
\equiv t_{DC}^0z^{-9/2}$ is a characteristic time scale for
double-Compton scattering, where $Y_p$ is the primordial helium mass
fraction, $\Omega_b$ is the fractional contribution of baryons to the
critical density, and $h$ is the Hubble constant in units of $100$ km
s$^{-1}$ Mpc$^{-1}$.
The solution to Eq.~\ref{dmu} at low
redshifts is given by
\begin{equation}
\label{solution}
\mu (z)= 1.4\int_0^{t(z\ll z_{\mu})}{\rm d}t\, {Q_B\over \rho_{\gamma}}
\,{\rm exp}-\biggl({z\over z_{\mu}}\biggr)^{5/2}\, ,
\end{equation}
where $z_{\mu}=(4\tilde{t}_0/5t_{DC}^0)^{-2/5}$ denotes the characteristic
redshift for \lq\lq freeze-out\rq\rq\ from double-Compton scattering.
The quantity $\tilde{t}_0 = 2.39\times 10^{19}\, {\rm s}$ determines 
the time-redshift relationship
via $t=\tilde{t}_0/z^2$ for cosmic times well before matter-radiation
equality. The characteristic redshift equals  
$z_{\mu} = 2.49\times 10^6$
for $\Omega_b h^2 = 0.0125$, and $Y_p=0.24$. 

The energy dissipation rate may be computed as the sum of the
contribution from all different magnetic field Fourier modes
\begin{eqnarray}
\label{QB}
Q_B & = & {1\over 8\pi k_c^3}\int {\rm d}^3k\, {d|\tilde{b}_k|^2\over
dt} 
\\
& = & {1\over 8\pi k_c^3}\int {\rm d}^3k\,|\tilde{b}_k|^2\,\, (2\, {\rm
Im}\omega)\,
{\rm exp}\biggl(-2\int {\rm Im}\omega \, dt\biggr)\, .\nonumber
\end{eqnarray}
Note that, by employing $\tilde{b}$, the decrease in magnetic field
energy density due to the expansion of the universe, which is not of
dissipative character, does not contribute to Eq.~\ref{QB}. Similarly,
the real part of the frequency in Eq.~\ref{dispersion} does not change the
total energy density in magnetic fields and fluid flow. When
$|\tilde{b}_k|$ and ${\rm Im}\,\,\omega$ are inserted in Eq.~\ref{QB} one
obtains 
\begin{eqnarray}
\label{energy}
{Q_B\over\rho_{\gamma}} =
{1\over\rho_{\gamma}^0}{\tilde{B}^2\over 8\pi}{(n+3)\over k_c^{n+3}}
\int_0^{k_c}{\rm d}k{k^{n+4}\over n_e^0\sigma_{Th}}\nonumber \\
\times {1\over 5(1+z)}
{\rm exp}\biggl(-{2\over 15}{k^2\tilde{t}_0\over
n_e^0\sigma_{Th}}{1\over
z^3}\biggr)\, .
\end{eqnarray}
It remains to substitute Eq.~\ref{energy} in
Eq.~\ref{solution} and perform the double-integral. An analytic
solution may be obtained  for $k_c\gg k_D^0z_{\mu}^{3/2}$ with
$k_D^0= (15n_e^0\sigma_{Th}/\tilde{t}_0)^{1/2}$, valid when the cutoff in
the spectrum $k_c$ is at much larger wavevectors than the
characteristic wavevector damped by photon viscosity
at redshift $z_{\mu}$ (see below). The solution is then simply
\begin{equation}
\label{total}
\mu = K{B_0^2\over 8\pi \rho_{\gamma}^0}\biggl({k_D^0\over k_c}
z_{\mu}^{3/2}\biggr)^{(n+3)}\, ,
\end{equation}
with $K= 1.4\,\Gamma (n/2 +5/2)\,\Gamma (3n/5 + 9/5)\,2^{-(n+5)/2}\,$
$(6/5)\,(n+3)$ a numerical constant of order unity. For example,
$K= 2.09, 1.10, 0.77, 0.78$ for magnetic field spectral indices
$n=1, 0, -1, -2$, respectively.
The scale $k_D^0z_{\mu}^{3/2}$ has a simple interpretation. It is the
scale that is damped by one e-fold at redshift $z_{\mu}$.  For the
above values of $\Omega_bh^2$ and $Y_p$ the corresponding comoving
wavelength is $\lambda_D = (2\pi)/(k_D^0z_{\mu}^{3/2}) = 395 {\rm
pc}$. Note that the ratio of magnetic field energy density and CMB
energy density is given by $B_0^2/(8\pi \rho_{\gamma}^0) = 
10^{-4}(B_0/3.24\times 10^{-8} {\rm Gauss})^2$. 

The limit may be put into context by comparing the prediction of 
Eq.~\ref{total} with the present upper limit from COBE/FIRAS,
$|\mu |< 9\times 10^{-5}$ at (95\% confidence level)~\cite{FIRAS} for a given
realization of the magnetic field spectrum. For example, when the
spectral cutoff is taken at $k_c = \pi/ r_H^{\rm QCD}$ with
$r_H^{\rm QCD}\approx 0.9\, {\rm pc}(T_{\rm QCD}/100{\rm MeV})^{-1}$
the approximate comoving QCD-horizon, as well as under the assumption
that the magnetic field energy density is in approximate equipartition
with the radiation energy density (i.e. $B_0 = 3.24\times 10^{-6}\,$
Gauss), one derives $n\, {}^>_{\sim}\, -1.35$. Note that there are
no constraints on such magnetic field spectra from BBN, 
since a significant fraction of the initial magnetic 
field energy density is already dissipated by neutrino viscosity
before neutrino decoupling~\cite{remark}.
For comparison, a scenario which produces uncorrelated magnetic
dipole fields in different QCD horizon domains, 
has a super-horizon power spectrum tail with $n=0$~\cite{H83}, and is
unconstrained. Nevertheless, magnetic field generation mechanism with 
resultant spectral indices of
$n=-1$~\cite{V91} and $n=-2$~\cite{EO94} 
have also been proposed. Furthermore, it is possible that processes,
such as inverse cascades in a  turbulent primordial
fluid, may significantly increase an initially small 
coherence length of magnetic fields~\cite{cascade}. 

The derived limit may be restated in an alternative form, 
independent of the initial statistical magnetic field properties given
by the unknown $l_c$ and $n$.  
Primordial magnetic fields with comoving coherence length $\approx 400$pc
cannot have fields of strength ${}^>_{\sim}\, 3\times 10^{-8}\,$Gauss 
since they would yield chemical potential distortions 
$\mu \, {}^>_{\sim}\, 10^{-4}$.
This is due to a nonthermal energy injection
$\Delta \rho_B/\rho_{\gamma} \, {}^>_{\sim}\, 10^{-4}$ induced by 
magnetic energy dissipation at redshift $z\approx 2\times 10^6$. One may
derive a similar limit for magnetic field dissipation at redshifts 
$z \, {}^<_{\sim}\,
4\times 10^4$ which would yield Compton $y$ distortions in the CMB.
For a magnetic field strength of $\approx 3\times 10^{-8}\,$Gauss,  
magnetic fields damp in the photon diffusion limit above redshift
$z \, {}^>_{\sim}\, 2\times 10^4$ with maximum
damping scale $\approx 0.3$Mpc~\cite{JKO98}.
Some additional damping of even slightly larger scales $\approx 0.6$Mpc
may occur shortly before recombination in the radiation free-streaming limit.
Since a magnetic field strength of $B_0\approx 3\times 10^{-8}\,$Gauss 
corresponds to a fractional CMB energy density of $\approx 10^{-4}$,
magnetic fields on such coherence scales would produce $y$ distortions
above the observational upper limit.

In summary,
we have shown that spatially varying magnetic fields
existing prior to the epoch of recombination, may lead to observable
distortions in the CMB blackbody. Distortions may result from the
injection of non-thermal energy into the CMB due to the dissipation 
of magnetic field induced peculiar flows in the baryon-photon fluid.
Our main result is, that magnetic fields of strength ${}^>_{\sim}\,
3\times 10^{-8}\,$Gauss with comoving 
coherence length $\approx 400$pc, may not be
present at redshifts $z\approx 2\times 10^6$, since they would yield
chemical potential distortions $\mu \, {}^>_{\sim}\, 10^{-4}$.
Similarly, magnetic fields of ${}^>_{\sim}\, 3\times 10^{-8}\,$Gauss on
scales $\sim 0.6$ Mpc are  not allowed since they would induce $y$ 
distortions shortly before recombination.
Limits on the spectral index of a stochastic magnetic field
distribution, depending on the total energy density in the field and 
its minimum coherence length, are also given.  
These limits are complimentary to limits on homogeneous primordial
magnetic field ($B_0\, {}^<_{\sim}\, 3\times 10^{-9}\,$Gauss) and limits on 
larger scales discussed above. Our limits are stronger than the present
Faraday rotation upper limits on these scales
which for  $\sim 400$ pc is only  $B_0\, {}^<_{\sim}\, 7\times 10^{-7}\,$
Gauss~\cite{private}.

\vskip 0.15in
We thank  Peter Coles and Rashid Sunyaev for many discussions and the
US-German program where this work was completed. This research was
supported in part 
at the University of Chicago by NSF through grant AST 94-20759 
and DOE grant DE-FG0291  ER40606.


\begin{references}
\bibitem{H83}C.~J.~Hogan, Phys.~Rev.~Lett. {\bf 51}, 1488 (1983).
\bibitem{fieldcreation} E.~R.~Harrison, MNRAS {\bf 147}, 279 (1970);
E. R. Harrison, MNRAS {\bf 165}, 185 (1973);  
M. S. Turner and L. M. Widrow, Phys. Rev. D {\bf 30}, 2743 (1988); 
J. Quashnock, A. Loeb, and D. N. Spergel, ApJ {\bf 344}, L49 (1989); 
R. H. Brandenberger, A.-C. Davis, A. M. Matheson, and M. Trodden, Phys. Lett. 
B {\bf 293}, 287 (1992); 
W. D. Garretson, G. B. Field, and S. M. Carroll, Phys. Rev. D {\bf 46}, 5346
 (1992); 
B. Ratra, ApJ {\bf 391}, L1 (1992); 
B. Ratra, Phys. Rev. D {\bf 45}, 1913 (1992); 
A. D. Dolgov, Phys. Rev. D {\bf} 48, 2499 (1993); 
A. D. Dolgov and J. Silk, Phys. Rev. D {\bf 47}, 3144 (1993); 
B. Cheng and A. V. Olinto, Phys. Rev. D {\bf 50}, 2421 (1994);  
A. P. Martin and A.-C. Davis, Phys. Lett. B {\bf 360}, 71 (1995); 
T. W. B. Kibble and A. Vilenkin, Phys. Rev. D {\bf 52}, 679 (1995);
M. Gasperini, M. Giovannini, and G. Veneziano, Phys. Rev. Lett. {\bf 75}, 
3796 (1995); 
D. Lemoine and M. Lemoine, Phys. Rev. D {\bf 52}, 1955 (1995);
G. Baym, D. B\"odecker, and L. McLerran, Phys. Rev. D {\bf 53}, 662 (1996);
G.~Sigl, A.~V.~Olinto, and K.~Jedamzik, Phys.~Rev. D {\bf 55} 4582 (1997).


\bibitem{V91}
T. Vaschaspati, Phys.~Lett.~B {\bf 265}, 258 (1991).

\bibitem{EO94} K. Enqvist, P. Olesen, Phys.~Lett.~B {\bf 329}, 195
(1994).

\bibitem{cascade} A.~Brandenburg, K.~Enqvist, and
P.~Olesen, Phys.~Lett.~B {\bf 392} 395 (1997);
T.~Shiromizu, Phys.~Lett.~B {\bf 443} 127 (1998); 
D.~T.~Son, Phys.~Rev.~D {\bf 59} 063008 (1999);
G.~B.~Field and S.~M.~Carroll,
astro-ph/9811206; 
M.~Christensson and M.~Hindmarsh, Phys.~Rev.~D {\bf 60} 063001 (1999).

\bibitem{BBN} B. Cheng, A. V. Olinto, D. Schramm, and J. Truran,
Phys. Rev. D {\bf 54}, 4714 (1996);
P. Kernan, G. Starkman, and T. Vachaspati, Phys. Rev. D 
{\bf 54}, 7202 (1996).

\bibitem{JKO98} K.~Jedamzik, V.~Katalini\'c, and A.~V.~Olinto,  
Phys.~Rev.~D {\bf 57}, 3264 (1998).

\bibitem{BFS97} J.~D.~Barrow, P.~G.~Ferreira, and J.~Silk,
Phys.~Rev.~Lett. {\bf 78}, 3610 (1997).

\bibitem{Fara} P.~Blasi, S.~Burles, and A.~V.~Olinto,
Ap. J. Letters, 512,
L79 (1999).


\bibitem{Pen95DFK99} P. Chen, Phys. Rev. Lett. {\bf 74}, 634
(1995); R.~Durrer, P.~G.~Ferreira, and T.~Kahniashvili, astro-ph/9911040.

\bibitem{SB98}
K.~Subramanian and J.~D.~Barrow,
Phys.~Rev.~Lett. {\bf 81}, 3575 (1998).

\bibitem{DKY98} R.~Durrer, T.~Kahniashvili, and A.~Yates, 
Phys.~Rev. D {\bf 58}, 123004 (1998).

\bibitem{HHZ97} A.~Loeb \& A.~Kosowsky, Astrophys.~J. {\bf 469}, 1 (1996);
D.~D.~Harari, J.~D.~Hayward, M.~Zaldarriaga,
Phys.~Rev. D {\bf 55}, 1841 (1997).

\bibitem{ADGR96}
J.~Adams, U.~H.~Danielsson, D.~Grasso, and H.~Rubinstein,
Phys.~Lett. B {\bf 388}, 253 (1996).

\bibitem{SB98_2}
K.~Subramanian and J.~D.~Barrow,
Phys.~Rev.~D {\bf 5808}, 3502 (1998).

\bibitem{distort} R.~A.~Sunyaev and Ya.~B.~Zel'dovich, Astrophys.  
Space Sci. {\bf 7}, 20 (1970);
R.~A.~Sunyaev and Ya.~B.~Zel'dovich,
Ann.~Rev.~Astron.~Astrophys. {\bf 18}, 537 (1980);
C.~Burigana, G.~De Zotti, and L.~Danese, Astr. \& Astrtophys.
{\bf 303}, 323 (1995).

\bibitem{FIRAS} D.~J.~Fixsen, E.~S.~Cheng, J.~M.~Gales, J.~C.~Mather, 
R.~A.~Shafer, and E.~L.~Wright,
Astrphys.~J. {\bf 473} 576 (1996). 

\bibitem{ZIS72} Ya.~B.~Zel'dovich, A.~F.~Illarionov, and
R.~A.~Sunyaev, JETP {\bf 35}, 643 (1972).

\bibitem{HS93} W.~Hu and J.~Silk, Phys.~Rev.~D {\bf 48}, 485 (1993).

\bibitem{PP99}
D.~Puy and P.~Peter, Int.~J.~Theor.~Phys. {\bf 38}, 205 (1999). 

\bibitem{remark} The magnetic field strength dependent
damping scale before neutrino decoupling is about $\sim 50$pc for
$B_0 \approx 3\times 10^{-6}\,$Gauss, but only $\sim 0.05$pc for
$B_0 \approx 3\times 10^{-8}\,$Gauss.

\bibitem{private} P.~Blasi, private communication.

\end{references}
\end{document}